# Validation of Golden Gate assemblies using highly multiplexed Nanopore amplicon sequencing

Adán A. Ramírez Rojas[1], Cedric K. Brinkmann[1] and Daniel Schindler[1,2,*]

[1] Max Planck Institute for Terrestrial Microbiology, Karl-von-Frisch-Str. 10, D-35043 Marburg, Germany

[2] Center for Synthetic Microbiology (SYNMIKRO), Philipps-University Marburg, Karl-von-Frisch-Str. 14, D-35032 Marburg, Germany

* Correspondence to: daniel.schindler@mpi-marburg.mpg.de

**Abstract**

Golden Gate cloning has revolutionized synthetic biology. Its concept of modular, highly characterized libraries of parts that can be combined into higher order assemblies allows engineering principles to be applied to biological systems. The basic parts, typically stored in level 0 plasmids, are sequence validated by the method of choice and can be combined into higher order assemblies on demand. Higher order assemblies are typically transcriptional units, and multiple transcriptional units can be assembled into multi-gene constructs. Higher order Golden Gate assembly based on defined and validated parts usually does not introduce sequence changes. Therefore, simple validation of the assemblies, e.g. by colony PCR or restriction digest pattern analysis, is sufficient. However, in many experimental setups, researchers do not use defined parts, but rather part libraries, resulting in assemblies of high combinatorial complexity where sequencing again becomes mandatory. Here we present a detailed protocol for the use of a highly multiplexed dual barcode amplicon sequencing using the Nanopore sequencing platform for in-house sequence validation. The workflow, called DuBA.flow, is a start-to-finish procedure that provides all necessary steps from a single colony to the final easy-to-interpret sequencing report.

1. Introduction

Once *in vitro* DNA constructs have been produced using the method of choice, they are transformed into recipient cells. Recipients are usually special *Escherichia coli* cloning strains for the propagation and archiving of DNA assemblies in the form of plasmids. After transformation, each colony typically contains a unique plasmid. Several methods have been established to confirm the integrity of plasmids. One rapid method is diagnostic PCR directly on cell material, called colony PCR (cPCR), which is suitable for selecting candidates for downstream analysis or applications, but does not confirm DNA sequence integrity [1]. An alternative method suitable for identifying candidates with similar results is diagnostic restriction digest pattern analysis of purified plasmid DNA [2]. However, the final step is the determination of the DNA sequence. The method of choice for sequencing plasmid DNA or amplicons is Sanger sequencing. Sanger sequencing allows the sequencing of stretches of approximately 1,000 base pairs (bp) [3]. Sanger sequencing provides reliable results in the laboratory routine, but becomes impractical when longer sequences or large numbers of constructs need to be validated.

In recent years, alternative sequencing methods have emerged that allow the parallel sequencing of DNA, the so-called Next Generation Sequencing (NGS) platforms [3]. This technology is a game changer, allowing the routine analysis of any type of DNA [4]. In addition to their primary application for whole genome sequencing and various other global sequence analysis methods, NGS platforms have been used to sequence plasmid DNA [5,6]. This approach allows the sequencing of a large number of plasmids in parallel. However, it requires either large investments in infrastructure or long turnaround times if external services have to be used. Furthermore, the analysis of the rather short reads is not as trivial as the analysis of Sanger sequencing results. In particular, structural variations may not be detected.

Based on the successes and limitations of NGS strategies, long-read sequencing technologies have emerged. Long-read sequencing technologies combine the advantage of long sequences, similar to Sanger sequencing, with the parallelization of NGS technologies [3]. Therefore, long-read sequencing technologies are often referred to as third-generation sequencing (TGS) technologies. Initially, TGS provided lower read quality with relatively high error rates. However, continuous improvements have made TGS technologies an established methodology. One notable long-read sequencing technology is based on the alteration of a current as DNA moves through a membrane inserted protein pore [7]. The current changes are specific to each base and allow a DNA strand to be decoded based on electrical signals. This technology is called nanopore sequencing and has become a widely used technology. Its advantage is the rather small infrastructure investment and its easy operation. Several protocols have been developed to use nanopore sequencing for the validation of DNA assemblies and whole plasmids [8,9] and commercial services are becoming available. However, in-house validation of DNA assemblies using nanopore sequencing is highly economical, especially when large numbers of DNA constructs need to be validated frequently.

We have recently developed a dual barcoding amplicon sequencing workflow called DuBA.flow which is available at GitHub (Fig. 1, https://github.com/RGSchindler/DuBA.flow) [10]. DuBA.flow is a start-to-finish workflow for generating and analyzing highly multiplexed amplicons, and returns easy-to-interpret result files, making the analysis of Nanopore sequencing results as easy as the analysis of Sanger sequencing data. Here we provide a detailed protocol for running our DuBA.flow pipeline from selected *E. coli* candidates to easy-to-interpret result files. The entire workflow can be performed within three days and is highly scalable. The protocol can be performed manually, but we also provide guidance on how to implement different types of laboratory automation equipment, making our DuBA.flow pipeline highly scalable. In particular, the use of acoustic dispensers allows for economic downscaling of reactions, making it highly economical.

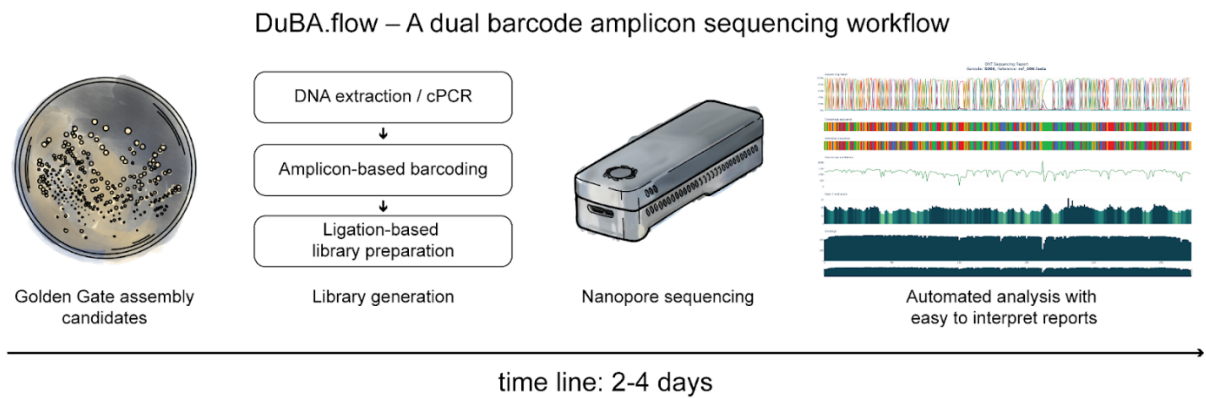

**Fig. 1 | DuBA.flow: a start to end long-read sequencing validation pipeline for assembled DNA constructs.** The pipeline starts with candidates obtained from the Golden Gate reaction transformation. Within the pipeline different strategies can be used to obtain barcoded amplicons which are handed over to the Ligation Sequencing library preparation and subsequent Nanopore sequencing. Obtained data can be passed on to the automated computational analysis pipeline which returns user-friendly files for sequence validation.

2.  **Materials**

The conduction of the described protocol requires the following non-standard and standard laboratory equipment and consumables.

1. MinION Nanopore sequencer and Flongle adapter are used in this protocol (*see* **Note 1**)
2. Reagents and standard consumables to conduct the Oxford Nanopore Technologies Ligation Sequencing Kit protocol (*see* **Note 2**)
3. Standard reagents, consumables and instrumentation for PCR reactions
4. Standard reagents and equipment for gel electrophoresis
5. Standard reagents, consumables and instrumentation for microbial culturing
6. Standard reagents, consumables and instrumentation for plasmid extraction and DNA clean up (*see* **Note 3**)
7. Standard reagents, consumables and instrumentation to determine DNA quality and concentration by spectrophotometric and fluorescent dye intercalation measurement
8. Standard micropipettes and consumables, 12 channel pipettes are recommended
9. Standard reagents, consumables and instrumentation for next-generation sequencing library preparation (e.g., magnetic racks, centrifuge, lo-bind DNA 1.5 mL reaction tubes, magnetic beads)
10. For downscaled and high-throughput cPCR workflows respective equipment and consumables are necessary (*see* **Note 4**)
11. Standard Thermomixer for 1.5 mL reaction tubes

*2.1 Computational requirements (Hardware & Software)*
Below are settings of hardware and software listed used in this chapter - they do not represent the minimum specifications necessary. If new equipment is going to be purchased it is important to check for the most recent hardware requirements for the used software and to operate the MinION Nanopore sequencer (*see* **Note 5, 6**).

1. Nanopore sequencing: Intel Core i7-10610U CPU, SDD: 1 TB, RAM: 32 GB (*see* **Note 7**)
   Operating System: Microsoft Windows 10 Enterprise
   Software: MinKNOW (up to version 23.11.4)
2. Basecalling: Intel Core i9-11900K CPU, SDDs: 1 TB, RAM: 64 GB, GPU: Nvidia GeForce RTX3090, 24 GB RAM, Compute Capability 8.6 (*see* **Note 8**)
   Operating System: Ubuntu Linux 22.04.3
   Software: CUDA (version 11.5), Guppy (up to version 6.5.7)
3. DuBA.flow analysis: Intel Core i7-10610U CPU, SDD: 1 TB, RAM: 32 GB (*see* **Note 9**)
   Operating System: Microsoft Windows 10 Enterprise

Software: Docker Desktop (version 4.12.0 (85629)), web browser (e.g., Mozilla Firefox, Google Chrome)

## 2.2 DNA oligonucleotides

Relevant oligonucleotides for the conduction of the protocol are provided in Table 1. Example primers attaching the barcodes are provided in Table 2. 100 µM stocks are generated with ddH$_2$O and stored at -20°C except indicated differently. For PCR and sequencing reactions 10 µM working stocks are generated with ddH$_2$O and stored at -20°C. **Important:** Oligonucleotides for transposase loading are solved in annealing buffer (50 mM NaCl, 40 mM Tris-HCl pH 8.0) to a stock concentration of 140 µM.

**Tab. 1 | Oligonucleotides for the conduction of the described protocol. Barcoding primers are listed in Table 2.**

| Name | Sequence (5´-3´)[a] | Information |
|---|---|---|
| SLo0100[b] | 5'[phos]CTGTCTCTTATACACATCT | Tn5-ME reverse for transposase loading [11] |
| SLo0151 | CCCAGTCACGACGTTGTAAAACG | M13 forward primer serving as control primer |
| SLo0152 | AGCGGATAACAATTTCACACAGG | M13 reverse primer serving as control primer |
| SLo0673[b] | **CCCAGTCACGACGTTGTAAAACG**AGATGTGTATAAGAGACAG | Linker Tn5-ME with added M13 forward sequence for annealing with SLo0100 |
| SLo0674[b] | **AGCGGATAACAATTTCACACAGG**AGATGTGTATAAGAGACAG | Linker Tn5-ME with added M13 reverse sequence for annealing with SLo0100 |
| SLo1577 | **CCCAGTCACGACGTTGTAAAACG**CGTCAATTGTCTGATTCGTTACCA | Forward primer for cPCR amplification with added M13 forward sequence |
| SLo1578 | **AGCGGATAACAATTTCACACAGG**CTTCTCTCATCCGCCAAAACA | Reverse primer for cPCR amplification with added M13 reverse sequence |

[a] bold letters indicate the overlaps for amplification (*see* **Note 10**) for the second PCR attaching the barcodes.
[b] Indicated oligonucleotides for Tn5 loading are solved in annealing buffer (50 mM NaCl, 40 mM Tris-HCl pH 8.0) to a stock concentration of 140 µM.

**Tab. 2 | Oligonucleotides for barcode attachment.[a]**

| Name | Sequence (5´-3´)[b] | Information |
|---|---|---|
| SLo2420 | **AAGAAAGTTGTCGGTGTCTTTGTG**CCCAGTCACGACGTTGTAAAACG | M13 forward primer with example barcode attached. |
| SLo2421 | **TCGATTCCGTTTGTAGTCGTCTGT**CCCAGTCACGACGTTGTAAAACG | M13 forward primer with example barcode attached. |
| SLo2422 | **GAGTCTTGTGTCCCAGTTACCAGG**CCCAGTCACGACGTTGTAAAACG | M13 forward primer with example barcode attached. |
| SLo2423 | **TTCGGATTCTATCGTGTTTCCCTA**CCCAGTCACGACGTTGTAAAACG | M13 forward primer with example barcode attached. |
| SLo2424 | **CTTGTCCAGGGTTTGTGTAACCTT**CCCAGTCACGACGTTGTAAAACG | M13 forward primer with example barcode attached. |
| SLo2516 | **AAGAAAGTTGTCGGTGTCTTTGTG**AGCGGATAACAATTTCACACAGG | M13 reverse primer with example barcode attached. |

| Name | Sequence (5´-3´)[b] | Information |
|------|---------------------|-------------|
| SLo2517 | **TCGATTCCGTTTGTAGTCGTCTGT**AGCGGATAACAATTTCACACAGG | M13 reverse primer with example barcode attached. |
| SLo2518 | **GAGTCTTGTGTCCCAGTTACCAGG**AGCGGATAACAATTTCACACAGG | M13 reverse primer with example barcode attached. |
| SLo2519 | **TTCGGATTCTATCGTGTTTCCCTA**AGCGGATAACAATTTCACACAGG | M13 reverse primer with example barcode attached. |
| SLo2520 | **CTTGTCCAGGGTTTGTGTAACCTT**AGCGGATAACAATTTCACACAGG | M13 reverse primer with example barcode attached. |

[a] five forward and reverse primer examples are provided. The barcodes can be freely defined by the user. Oxford Nanopore Technologies provides the sequences for their 96 established barcodes.
[b] bold letters indicate the barcode sequence.

## *2.3 Enzymes*

Any enzyme with corresponding properties can be used. For the presented protocol used enzymes were purchased from New England Biolabs (NEB) except for the Tn5 transposase and the HiFi polymerase.

1. KAPA HiFi HotStart ReadyMix (*see* **Note 11**).
2. Transposase; here in-house purified Tn5 transposase [12,11]
3. Proof-reading polymerase
4. Recommended enzymes for the Oxford Nanopore Technologies Ligation Sequencing kit enzymes (*see* **Note 12**)

## *2.4 Chemicals, buffers and media components*

1. Ethanol (*see* **Note 13**)
2. Magnetic beads (*see* **Note 14**)
3. DNA intercalating dye for DNA visualization after gel electrophoresis (*see* **Note 15**)
4. Annealing buffer: 50 mM NaCl, 40 mM Tris-HCl pH 8.0
5. Tn5 dilution buffer: 10 mM Tris-HCl pH 7.5, 150 mM NaCl
6. 4X Tagmentation buffer: 40 mM Tris-HCl pH 7.5, 40 mM MgCl2
7. 100% Dimethylformamide (DMF)
8. 100% Dimethyl sulfoxide (DMSO)
9. Glycerol

## *2.5 Consumables*

1. PCR reaction vessels from single tubes to 384-well plates
2. Appropriate seal for PCR plates

## 3. Methods

### 3.1 Generation of barcoded amplicons

The key element of the described dual barcoding procedure is that the initial PCR or enzymatic fragmentation is performed with oligonucleotides containing a landing sequence for a subsequent PCR to add the user defined barcodes (Fig. 2). This strategy has the advantage that the primers containing the barcodes are highly flexible. By simply changing the initial PCR primers to a new target sequence, the barcode primers can be reused. The protocol describes three different approaches to generate the dual barcoded amplicons. The three different approaches differ only in the initial step (Fig. 2B-D). Enzymatic fragmentation has the advantage of providing dual barcoded amplicons for global analysis beyond a specific amplicon. However, this strategy results in a reduced sequencing output due to the generation of fragments with the same barcode at both ends (Fig. 2D). The generation of dual barcoded amplicons can be performed within half a day from purified DNA or single colonies.

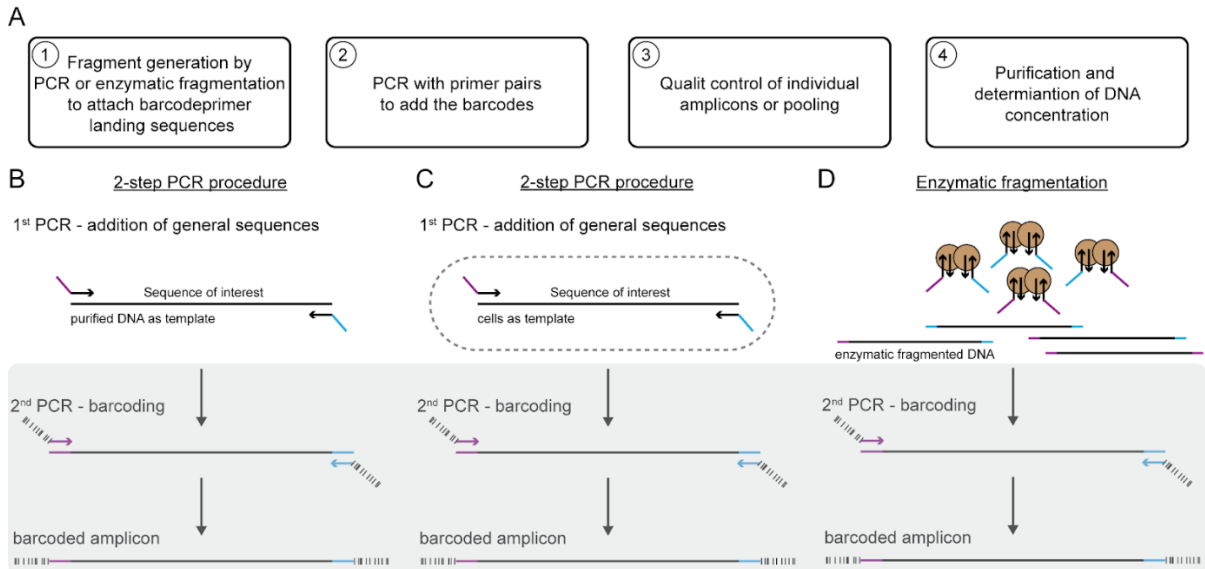

**Fig. 2 | Workflow and concept of the different types of dual barcoded DNA fragment generation. (A)** Workflow for the generation of the dual barcoded amplicons. **(B) (C)** and **(D)** visualize the different strategies used to generate DNA fragments with the landing sequences for the primers attaching the barcode. The PCR attaching the barcodes is in all scenarios identical indicated by the gray background. The strategies differ in using purified DNA as template for a distinct sequence of interest (B), using cells as template (indicated by dotted rod shaped bacterial cell) for a distinct sequence of interest (C) or enzymatically fragment purified DNA for whole plasmid analysis (D). The strategies were developed in parallel to allow reuse of the barcoding primers for any type of application by only swapping the amplification primers with the landing sequences for in the first step of B and C.

### 3.1.1 Generation of barcoded amplicons from purified plasmids

1. Design primer sequences for target plasmid to generate PCR amplicons with M13 barcode primer landing sequences (Fig. 2B)(*see* **Note 16**). Below is an example design, in this case "N" indicates the amplification specific sequence to be added and the determined sequence are the M13 forward and reverse sequences:

    forward primer    5´-**CCCAGTCACGACGTTGTAAAACG**NNNNNNNNNNNNNNNNNNNN-3´
    reverse primer    5´-**AGCGGATAACAATTTCACACAGG**NNNNNNNNNNNNNNNNNNNN-3´

    **Important:** These primers are used for all parts of this protocol where an initial PCR amplification is required (all except the Tn5 procedure) to add the barcode primer landing sequences.

2. Extract the plasmids (or genomic DNA if applicable [*see* **Note 17**]) using the method of choice and perform standard quality control steps.

3. Prepare a 10 µl PCR reaction mixture using a proofreading DNA polymerase (*see* **Note 18, 19**). In this case, the example primer pair used is SLo1577/SLo1578, which amplifies inserts in level 0 plasmids from the in-house used Golden Gate vectors [13,14,10]. The reaction mix is as follows:

| Reagent | Stock concentration | Volume (*see* Note 19) |
|---|---|---|
| **Reaction buffer** | 5X | 2 µL |
| **dNTPs** | 10 mM | 0.2 µL |
| **forward primer (SLo1577)** | 2 µM | 0.625 µL |
| **reverse primer (SLo1578)** | 2 µM | 0.625 µL |
| **DNA template** | 1-10 ng | - |
| **Polymerase** | 2 U/µL | 0.1 µL |
| **ddH$_2$O** | - | to 10 µL |

4. Perform PCR reaction in a thermal cycler using the following settings (*see* **Note 20**):

| Step | Settings | Cycles |
|---|---|---|
| **Initial denaturation** | 98°C for 20 sec | - |
| **PCR** | 98°C for 20 sec<br>66°C for 20 sec<br>72°C for 30 sec | 10 x |
| **Final extension** | 72°C for 1 min | - |
| **Hold** | 12°C | - |

5. Prepare a 1:10 dilution of the initial PCR to be used as a template for the subsequent PCR.
6. Prepare 10 µl PCR reactions using unique combinations of barcode primer pairs (examples are provided in Table 2):

| Reagent | Stock concentration | Volume (*see* Note 19) |
|---|---|---|
| **Reaction buffer** | 5X | 2 µL |
| **dNTPs** | 10 mM | 0.2 µL |
| **forward primer** | 2 µM | 1 µL |
| **reverse primer** | 2 µM | 1 µL |
| **DNA template** | - | 1 µL |
| **Polymerase** | 2 U/µL | 0.1 µL |
| **ddH₂O** | - | to 10 µL |

7. Perform the PCR reaction in a thermal cycler using the following settings:

| Step | Settings | Cycles |
|---|---|---|
| **Initial denaturation** | 98°C for 30 sec | - |
| **PCR** | 98°C for 10 sec<br>66°C for 10 sec<br>72°C for 30 sec | 25 x |
| **Final extension** | 72°C for 1 min | - |
| **Hold** | 12°C | - |

8. Pool all PCR reactions and perform PCR purification using the method of choice (*see* **Note 3, 21**).
9. Determine DNA concentration using a spectrophotometer and a fluorescent dye-based method (*see* **Note 22**).
10. It is recommended to perform an electrophoretic analysis of the DNA to ensure that the size distribution of the DNA represents the expected sizes.
11. Store samples at -20°C until further use or proceed directly to step 3.3 (*see* **Note 23**).

### *3.1.2 Generation of barcoded amplicons from microbial colonies*

1. The protocol has been tested for *E. coli* and *S. cerevisiae* colonies (*see* **Note 24**). The cPCR procedure is described in the next two points, respectively.
2. For *E. coli* candidates: Transfer cell material to a PCR tube (e.g., using a sterile toothpick) containing 50 µL ddH₂O and vortex briefly mix to resuspend. Use 1 µL as template in the PCR reaction (step 4). Ensure that that cell material is preserved e.g. by storing the cell solution in the refrigerator or by cultivating it in appropriate media for subsequent cryopreservation.
3. For *S. cerevisiae* candidates: Transfer cell material to a PCR tube (e.g., using a sterile toothpick) containing 50 µL ddH₂O and vortex briefly to resuspend. Transfer 25 µL of the cell suspension to a PCR tube containing 25 µL of 40 mM NaOH. Store the remaining 25 µL cell suspension in the refrigerator. Transfer the PCR tube with 50 µL suspension to a PCR cycler or heat block and lyse cells at 95°C for 15 min. Use 1 µL as template in the PCR reaction.
4. Prepare PCR reactions with 10 µL total volume depending on the number of candidates according to the following reaction mix:

| Reagent | Stock concentration | Volume (*see* Note 19) |
|---|---|---|
| **Reaction buffer** | 5X | 2 µL |
| **dNTPs** | 10 mM | 0.2 µL |
| **forward primer (SLo1577)** | 2 µM | 0.625 µL |
| **reverse primer (SLo1578)** | 2 µM | 0.625 µL |
| **DNA template** | - | 1 µL |
| **Polymerase** | 2 U/µL | 0.1 µL |
| **ddH$_2$O** | - | to 10 µL |

5. Perform the PCR reaction in a thermal cycler using the following settings (*see* **Note 20**):

| Step | Settings | Cycles |
|---|---|---|
| **Initial denaturation** | 98°C for 3 min | - |
| **PCR** | 98°C for 20 sec<br>66°C for 20 sec<br>72°C for 30 sec | 10 x |
| **Final extension** | 72°C for 1 min | - |
| **Hold** | 12°C | - |

6. Dilute the PCR reaction 100-fold to serve as template for the next step to ensure maximum of barcoded amplicons.
7. Prepare PCR reactions with 10 µL total volume depending on the number of candidates according to the following reaction mix. Ensure that each reaction has a unique combination of barcode primer pairs (see Table 2 for examples):

| Reagent | Stock concentration | Volume (*see* Note 19) |
|---|---|---|
| **Reaction buffer** | 5X | 2 µL |
| **dNTPs** | 10 mM | 0.2 µL |
| **forward primer** | 2 µM | 1 µL |
| **reverse primer** | 2 µM | 1 µL |
| **DNA template** | - | 1 µL |
| **Polymerase** | 2 U/µL | 0.1 µL |
| **ddH$_2$O** | - | to 10 µL |

8. Perform the PCR reaction in a thermal cycler using the following settings:

| Step | Settings | Cycles |
|---|---|---|
| **Initial denaturation** | 98°C for 30 sec | - |
| **PCR** | 98°C for 10 sec<br>66°C for 10 sec<br>72°C for 30 sec | 25 x |
| **Final extension** | 72°C for 1 min | - |
| **Hold** | 12°C | - |

9. Pool all PCR reactions and perform PCR purification using the method of choice (*see* **Note 3, 21**).
10. Determine DNA concentration using a spectrophotometer and a fluorescent dye-based method (*see* **Note 22**).
11. It is recommended to perform an electrophoretic analysis of the DNA to ensure that the size distribution of the DNA corresponds to the expected sizes.
12. Store samples at -20°C until further use or proceed directly to step 3.3 (*see* **Note 23**).

### *3.1.3 Generation of barcoded amplicons from plasmids by enzymatic fragmentation*

Prelude: The following protocol is based on the use of the Tn5$_{(R27S, E54K, L372P)}$ transposase developed by Hennig *et al*. [11]. This mutated version of Tn5, produced in-house according to the author´s procedures, was selected because it generates larger fragments than the regular Tn5 used for library preparation for short-read sequencing technologies. The DNA fragmentation and adapter addition steps are based on the procedure described by Vonesch *et al*. [12]. Limitations of this procedure are that the transposase integrates adapter sequences in a random manner. This results in fragments with the same barcode at both ends of the sequence. In the theoretical ideal scenario, 50% of the fragments would have two different barcodes at both ends and 25% would have identical barcodes of either the forward or reverse landing pad sequence.

1. Extract the plasmids of interest using the method of choice and perform quality control by determining the A$_{260/280}$ using a spectrophotometer and determine the DNA concentration using a fluorescence assay of choice.
2. Dilute samples to a final concentration of 10-20 ng/µL and store at -20°C until use.
3. Prepare annealed oligonucleotides for Tn5 loading. Resuspend lyophilized oligonucleotides SLo0100, SLo0673 and SLo0674 (Table 2) in annealing buffer to a stock concentration of 140 µM and mix independently: add one volume (50 µL) of SLo0673 or SLo0674 to one volume of SLo0100. The resulting working stock is 70 µM.
4. Perform the following program in a thermal cycler to anneal the oligonucleotides:

| Step | Settings | Time |
|---|---|---|
| **Denaturation** | 95°C | 5 min |
| **Annealing**[a] | 95 to 65°C | 0.1°C/sec |
| **Hold** | 65°C | 5 min |
| **Slow cooling**[a] | 65 to 4°C | 0.1°C/sec |
| **Hold** | 4°C | - |

[a] This can usually be programmed by setting the PCR ramp rate to the minimum value.

5. Dispense the annealing reaction in 10 µL aliquots on ice into PCR tubes and store at -20°C.
6. Prepare the following reaction mix to load the Tn5 with the adapters:

| Reagent | Stock concentration | Volume |
|---|---|---|
| **Reaction buffer** | 0.5 mg/mL | 2 µL |
| **Annealed adapter (M13 forward)** | 140 µM | 1 µL |
| **Annealed adapter (M13 reverse)** | 140 µM | 1 µL |
| **Tris-HCl pH 7.5** | 20 mM | 8 µL |
| **Total** | - | 11 µL |

7. Incubate at 23°C for 30-60 min at 300 rpm in a thermomixer (*see* **Note 25**).
8. Add 89 µL dilution buffer for immediate use. Alternatively, add 11 µL of 100% glycerol to store the adapter-loaded Tn5 mixture at -20°C (50% final concentration). For the stored Tn5 mixture, dilute 1:5 with dilution buffer prior use.
9. Perform enzymatic fragmentation of each plasmid DNA separately by preparing the reaction mixture listed below. Keep all the components and final reactions on ice.

| Component | Stock concentration | Volume |
|---|---|---|
| **Tagmentation buffer** | 4X | 1.25 µL |
| **DMF** | 100% | 1.25 µL |
| **Diluted Tn5 adapter complex** | - | 1.25 µL |
| **Plasmid DNA (10-20 ng/µL)** | - | 1.25 µL |
| **Total** | - | 5 µL |

10. Incubate the reactions in a thermal cycler at 55°C for 30 sec and inactivate at 80°C for 5 min.
11. Store samples on ice if barcoding PCR is to be performed subsequently. Alternatively, store at -20°C until further use.
12. Prepare 7 µl PCR reactions using KAPA HiFi HotStart ReadyMix (Roche) to add barcodes to the enzymatically fragmented DNA (*see* **Note 26**). Add a unique barcode combination to each reaction using different primer pairs (see Table 2 for examples of primers) (*see* **Note Y**):

| Reagent | Stock concentration | Volume (*see* Note 19) |
|---|---|---|
| **KAPA HiFi Master Mix** | 2X | 3.5 µL |
| **DMSO** | 100% | 0.52 µL |
| **forward primer** | 10 µM | 0.21 µL |
| **reverse primer** | 10 µM | 0.21 µL |
| **DNA template** | - | 2 µL |
| **ddH₂O** | - | to 7 µL |

13. Perform the following PCR reaction in a thermal cycler which is optimized for the M13 landing sequences (*see* **Note 20**):

| Step | Settings | Cycles |
|---|---|---|
| **Gap filling (*see* Note 27)** | 72°C for 3 min | |
| **Initial denaturation** | 95°C for 30 sec | - |
| **PCR** | 98°C for 20 sec<br>66°C for 15 sec<br>72°C for 5 min | 20 x |
| **Final extension** | 72°C for 10 min | - |
| **Hold** | 12°C | - |

14. Pool all PCR reactions and perform PCR purification using the method of choice (*see* **Note 3, 21**).
15. Determine DNA concentration using a spectrophotometer and a fluorescent dye-based method (*see* **Note 22**).
16. Perform an electrophoretic analysis of the DNA to ensure the distribution of DNA sizes represents the expected sizes. In this case, a DNA smear and no distinct band is expected.
17. Store samples at -20°C until further use or proceed directly to step 3.3 (*see* **Note 23**).

### 3.2   *Utilizing automation for downscaling of reactions*

High-throughput library preparation requires a large amounts of reagents and is prone to human error. Recent technological developments allow for the use of laboratory automation to increase the number of samples while reducing the use of reagents. Acoustic dispensers have become a valuable tool for dispensing of complex reaction mixtures [15,16]. Acoustic dispensers are used to automate and parallelize Golden Gate DNA assembly reactions [13]. Acoustic dispensers have previously been used to scale down the preparation of short-read sequencing libraries for plasmid validation [5,6]. The acoustic dispenser is used here to accurately dispense the barcode primer pair combinations. The experimental procedure takes about half a day including PCR reactions and DNA purification. This procedure has been tested to sequence up to 1536 amplicons on a single Flongle flow cell [10].

### *3.2.1 Barcoded amplicons from purified plasmids*

1. Purify plasmids in 96-well format using method of choice. Alternatively, arrange plasmids in 96-well format.
2. Prepare a 1:100 plasmid dilution and transfer 1 µL per sample into a 384 well plate using 8-, 12-, or 96-channel pipettes (*see* **Note 28**).
3. Dispense 1 µL of the first PCR master mix into each well. The reaction mixture per sample is as follows:

| Reagent | Stock concentration | Volume (*see* **Note 19**) |
| --- | --- | --- |
| **Reaction buffer** | 5X | 0.4 µL |
| **dNTPs** | 10 mM | 0.04 µL |
| **forward primer** | 2 µM | 0.125 µL |
| **reverse primer** | 2 µM | 0.125 µL |
| **Polymerase** | 2 U/µL | 0.02 µL |
| **ddH$_2$O** | - | 0.29 µL |

4. Seal the PCR plate with an appropriate seal.
5. Ensure that the reaction mix is at the bottom of all wells by gently spinning the plate in an appropriate centrifuge.
6. Perform the first PCR with optimized settings for the used primer pairs used and the expected amplicon size. For the standard primer pair used in this study (SLo1577/SLo1578) and expected amplicon sizes the settings are as follows:

| Step | Settings | Cycles |
| --- | --- | --- |
| **Initial denaturation** | 98°C for 20 sec | - |
| **PCR** | 98°C for 20 sec<br>66°C for 20 sec<br>72°C for 30 sec | 10 x |
| **Final extension** | 72°C for 1 min | - |
| **Hold** | 12°C | - |

7. Remove the seal from the plate and dispense unique barcoded primer combinations into each reaction mix. In this protocol, an acoustic dispenser is used to dispense 100 nL of each oligonucleotide (final concentration of 0.2 µM) into the reaction mix in 2.5 or 25 nL increments, depending on the machine type (*see* **Note 29**).
8. Prepare the master mix for the second PCR reaction as described below and dispense 3 µL into each well. Total reaction volume is 5 µL.

| Reagent | Stock concentration | Volume (*see* Note 19) |
|---|---|---|
| **Reaction buffer** | 5X | 1 µL |
| **dNTPs** | 10 mM | 0.1 µL |
| **Polymerase** | 2 U/µL | 0.05 µL |
| **ddH$_2$O** | - | 1.85 µL |

9. Seal the PCR plate with an appropriate seal and perform the second PCR with optimized settings for the primer pairs. In the case of the standard M13 barcode primer pair used in this study (Table 2), the settings are as follows:

| Step | Settings | Cycles |
|---|---|---|
| **Initial denaturation** | 98°C for 30 sec | - |
| **PCR** | 98°C for 10 sec<br>66°C for 10 sec<br>72°C for 30 sec | 25 x |
| **Final extension** | 72°C for 1 min | - |
| **Hold** | 12°C | - |

10. Combine all PCR reactions in a single tube and proceed with preferred amplicon purification procedure (*see* **Note 30**).
11. After purification, determine DNA quality and concentration using a spectrophotometer and a fluorescence-based method.
12. Test the purified DNA by an electrophoresis method of choice.
13. Store samples at -20°C until further use or proceed directly to step 3.3 (*see* **Note 23**).

### 3.2.2 *Barcoded amplicons from microbial colonies*

1. Bacterial cells are arrayed in a 384 grid format on solid agar using a colony picking robot.
2. Solid plates are incubated at 37°C for several hours to obtain sufficient but not excessive cell material. Alternatively, liquid cultures can be grown overnight at 37°C with vigorous shaking. Here, a humidity-controlled incubator to incubate 384-well microtiter plates at 550 rpm, 3 mm shaking, 37°C, and 75% relative humidity is used. With these settings, it is possible to work without a seal for microbial culture (*see* **Note 31**). Liquid culture may have an advantage in that the transfer of cell material is more standardized. At the same time, however, transferring media to low-volume cPCR reactions can cause inhibition of the PCR reaction, especially if cultures are incubated too long.
3. Dispense 2 µL cPCR master mix to the wells of a 384 PCR plate. The reaction mixture is as follows:

| Reagent | Concentration | Volume (*see* Note 19) |
|---|---|---|
| **Reaction buffer** | 5X | 0.4 µL |
| **dNTPs** | 10 mM | 0.04 µL |
| **forward primer** | 2 µM | 0.125 µL |
| **reverse primer** | 2 µM | 0.125 µL |
| **Polymerase** | 2 U/µL | 0.02 µL |
| **ddH$_2$O** | - | 1.29 µL |

4. Transfer cell material using a 384-well pinpad either from the incubated solid media plate or the liquid culture to the cPCR PCR plate containing the dispensed cPCR master mix (*see* **Note 32**).
5. Ensure that the reaction mix is at the bottom of all wells by gently pulse-centrifugation of the plate. Too much centrifugation may cause cells to pellet at the bottom of the wells and prevent efficient cPCR (*see* **Note 33**).
6. From this step onwards the protocol is identical to 3.2.1 from step 6. Perform the following steps according to step 6 and following of 3.2.1.

### *3.3    Nanopore sequencing library preparation and Nanopore sequencing*

Nanopore sequencing library preparation relies on the library preparation kits provided by Oxford Nanopore Technologies. For the DuBA.flow workflow the Ligation Sequencing kit series is used (*see* **Note 34**). The procedure of the library preparation takes around two hours and sequencing on a Flongle flow cell, which is sufficient for most applications in regard to DNA assembly validation takes 24 hrs.

1. If not already done, determine the concentration of the sample using a fluorescence-based method. It is also recommended to measure the sample with a spectrophotometer to determine the A$_{260/280}$ ratio, which should be 1.8.
2. The procedure outlined is based on the Oxford Nanopore Technologies Ligation Sequencing Kit (SQK-LSK109). Due to the constantly improving chemistry, it is recommended to frequently check for updated protocols and to follow the latest procedure for the available kit version (*see* **Note 35**).
3. Perform library preparation according to the most recent protocol (*see* **Note 36**).
4. Assemble the Nanopore sequencer with the Flongle adapter and connect the sequencer to the operating computer.
5. Perform the sequencing experiment with the preset settings for Flongle flow cells. No live basecalling is performed in this scenario. If the used hardware supports GPU-based basecalling, live basecalling is recommended.

6. After the sequencing run, disassemble and store all equipment appropriately. If Flongle cells are used, dispose according to the local regulations. If standard flow cells are used act according to the current return procedure of Oxford Nanopore Technologies (*see* **Note 37**).
7. Appropriately store and backup the primary, not basecalled data.
8. Perform basecalling of raw data according to the preferred settings. In this protocol basecalling is performed using the Oxford Nanopore Technologies basecaller Guppy in the GPU mode (*see* **Note 38**). The following command is used in a Linux OS terminal:

```
guppy_basecaller -i /path_to_input_fast5_files/ -r -s
/path_to_output_folder/ --flowcell type_of_flow_cell --kit
type_of_library_kit --device cuda:0 --num_callers 24 --
disable_qscore_filtering --compress_fastq
```

**Important:** The elements in *italics* need to be adjusted by the operator.

Proceed to step 3.4 to analyze the sequencing data using the automated DuBa.flow analysis pipeline.

### 3.4    *Analysis pipeline and validation of sequences*

The most important part of sequence analysis is a rapid and throughout analysis pipeline and easy to interpret result files (Fig. 3). The DuBA.flow workflow provides a start-to-end solution for in-house sequence analysis. The computational analysis software is available at https://github.com/RGSchindler/DuBA.flow under a CC BY-NC-SA 4.0 license (*see* **Note 39**). The analysis pipeline can be executed on a standard computer and needs minimal computational resources and time. However, a bottleneck may be the basecalling of the raw data which requires a dedicated GPU and the respective computational environment for rapid base calling. The current basecaller of Oxford Nanopore Technologies is Guppy and is available from the Nanopore Community (*see* **Note 40**). The DuBA.flow analysis takes only little hands on and computation depends on the complexity and amount of data but usually the needed time is neglectable.

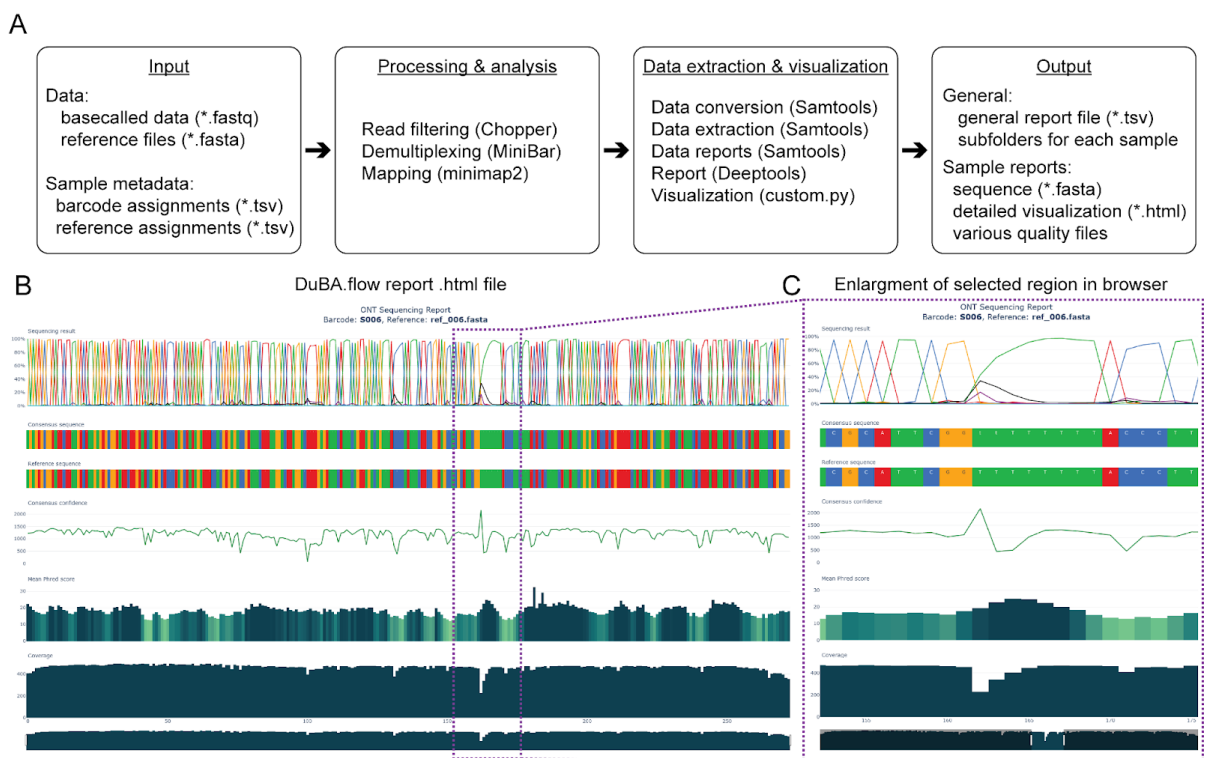

**Fig. 3 | DuBA.flow analysis pipeline and user-friendly output files. (A)** Workflow of the automated analysis pipeline which relies on user provided sequencing data, reference data and two sample metadata files. The data is processed automatically and returns user-friendly analysis files in subfolders for each sample, as well as an overview report. **(B)** Example of interactive results file generated by the analysis workflow. The .html file can be opened with standard web browsers and allows the user to move and zoom along the navigation bar; the indicated purple dotted box represents the enlargement shown in **(C)**. Zooming in allows to analyze the Nanopore sequencing results similar to Sanger sequencing traces. Figure and figure legend from DuBa.flow GitHub description (https://github.com/RGSchindler/DuBA.flow).

1. If no live basecalling was performed during the sequencing experiment the raw data has to be basecalled (see step 8 in 3.3).
2. The DuBA.flow computational pipeline is provided as Docker image (*see* **Note 41**). The most recent documentation can be found at https://github.com/RGSchindler/DuBA.flow.
3. Install Docker Desktop (https://www.docker.com/) on the machine used for analysis purposes if not already installed (*see* **Note 42, 43**).
4. After successfully installing and launching Docker, the analysis pipeline can be pulled using the command "`docker pull cedkb/dubaflow:0.4`" from the operating system terminal (*see* **Note 44**). This step is only necessary when (a) installing DuBA.flow for the first time or (b) after deleting the DuBA.flow image from Docker.
5. Before submitting the input files and running the analysis pipeline, it is essential to bind the volume of the Docker container to a specific folder on the host machine. To initiate the container and create the volume mount, type the following command in the system terminal:

```
"docker run --name dubaflow --workdir /home/dubaflow -it --
mount type=bind,src=home_path,dst=/home/dubaflow/input
cedkb/dubaflow:0.4"
```

In this command, replace "`home_path`" with the path to the folder that provides the input files on the host machine.

6. The following files must be provided to execute the computational analysis pipeline (*see* **Note 45**):

    (i) Provide reference files as separate, valid FASTA (`*.fa` and/or `*.fasta`) files within the "`references`" directory (*see* **Note 46**).
    (ii) For demultiplexing, DuBA.flow uses minibar [17] and requires the "`IndexCombination.tsv`" file, which conforms to the "`barcode demultiplex file format`" described in the minibar documentation.
    (iii) To associate samples with their respective references, a similar "`references.tsv`" file is used, which has two columns: SampleID and ReferenceFile, the latter being the name of the corresponding FASTA file.
    (iv) The base-called raw reads are provided in a single "`input.fastq.gz`" file.

7. The terminal, after executing the "`docker run`" command, should seamlessly transition to the terminal interface of the "`dubaflow`" docker container. After placing the input files in the designated folders, the pipeline can be easily initiated by running the command "`python3 DuBaFlow.py`" (*see* **Note 47**).
8. First DuBA.flow checks if all required files are available, if not an error message is returned, otherwise the calculation pipeline calculates everything automatically. Keep in mind the computation time needed depends on the specifications of the used computer and the complexity of the data.
9. Once the DuBa.flow analysis pipeline has been completed, an output directory is created in the specified home path, containing

    (i) A comprehensive summary report file
    (ii) Distinct subfolders for each sample encapsulating individual reports

A report including the processing status of each sample and provides a statistical breakdown highlighting metrics such as the total number of sequence reads, the number of reads successfully mapped, the observed error rate, the average read length, and the average quality score. The individual sample reports provide a range of valuable reports and output files, including the interactive easy-to-use HTML report file (Fig. 3B) that conveys all essential information and a consensus file in FASTA.

## 3.5 Summary & Perspectives

Golden Gate DNA assembly is growing in popularity, complexity, and throughput. Long-read sequencing has become a widely applicable technology for fast and accurate in-house sequence validation and genome sequencing. The step-by-step protocol outlined here provides a detailed description of how to use the DuBA.flow workflow for rapid analysis of Golden Gate DNA assemblies from low to high throughput. The user-friendly result files of the analysis pipeline are valuable for sequence validation of basic parts, transcription units and/or higher order Golden Gate assemblies.

## 4. Notes

1. Any other Oxford Nanopore Technologies Nanopore sequencer should be compatible with this procedure. If other manufacturers/sequencing technologies are used, the protocol must be adapted accordingly.
2. Other Nanopore sequencing library kits may be used. However, we experienced the best results with the ligation sequencing kit.
3. Any commercial kit or in-house procedure can be used. We used the open source magnetic bead protocols available at www.bomb.bio for plasmid extraction and DNA purification [18].
4. This protocol uses the Echo525/650T, the ARI Cobra, the Singer Instruments PIXL, and the Singer Instruments Rotor HDA+. Other instruments may be used. However, the procedure is optimized for the combination of instruments outlined.
5. Depending on the Nanopore sequencing platform used, a machine for operating the Nanopore sequencer may already be included (e.g. GridION).
6. The settings below are the systems used in this chapter; they are neither minimum nor maximum specifications. Other operating systems and software may be used.
7. It is recommended to check the latest hardware requirements for Nanopore sequencing before ordering a computer to operate a Nanopore sequencer.
8. GPU based basecalling requires an Nvidia GPU with a minimum Compute Capability. It is recommended to check Guppy (or alternative basecaller) specifications for the latest requirements.
9. Hardware requirements are subject to change. For the latest DuBA.flow analysis information, visit https://github.com/RGSchindler/DuBA.flow.
10. The overhangs can be freely defined by the user. For historical reasons, the M13 sequences are used in this application. The only requirements for the overhangs are that the forward and reverse are not identical and do not bind elsewhere in the DNA samples.
11. Other products may be suitable, but this reagent worked best in the setup described.
12. All enzymes and consumables were used according to the Oxford Nanopore Technologies Ligation Sequencing Kit protocol, with the exception of the FFPE enzyme mix, which was not used, and the AMPure XP beads, which were purchased from an alternative supplier.
13. Always prepare ethanol dilutions fresh to avoid reduced alcohol content due to evaporation.
14. Protocols are available to generate in-house replacements for commercial supplies. A good reference is e.g. https://dx.doi.org/10.17504/protocols.io.bkppkvmn.
15. Thiazole orange has been established as a reliable and inexpensive dye for staining DNA [19]. Thiazole Orange is dissolved in dimethyl sulfoxide (DMSO) (10,000 x stock concentration: 13 mg/mL).
16. The M13 sequences are used for historical reasons, but can be replaced by any other sequence. It can be an *in silico* designed sequence. These sequences should be unique to serve as ideal landing sequences for the barcoding primers.
17. This procedure can also be used to analyze amplicons from any other type of DNA, such as amplicons from genomes or metagenomic samples. Only the initial primers from step 1 need to be redesigned to amplify e.g. bacterial 16S or fungal ITS sequences.
18. We have obtained good results using the Q5 High-Fidelity DNA Polymerase (NEB) with the standard buffer. However, each new primer pair is tested with a gradient PCR for optimal amplification. The optimal primer concentration may vary with other polymerases.
19. It is recommended to prepare a master mix for the PCR reaction for the number of samples to be analyzed. When preparing a master mix, it is recommended to include a 10% volume overage to account for volume loss during dispensing. The master mix contains all elements except the DNA template and barcode attachment primer, which are unique to each sample. These reagents are added to the individual reactions after dispensing the Master Mix. When using small amounts of oligonucleotides, it is recommended to prepare 2 µM working solutions for more accurate pipetting.
20. If other primer sequences are used, it is recommended to test their optimal annealing temperature by performing a gradient PCR and gel electrophoresis to check for unspecific amplification.
21. When analyzing large numbers of amplicons, it does not make sense to purify and quality control each amplicon individually. It is more economical to repeat failed candidates in the next sequencing experiment than to perform quality control on individual amplicons. An appropriate number of candidates should be analyzed from the start, taking into account incorrect assemblies and amplification failures.
22. Spectrophotometric analysis is intended to provide information on DNA purity via the $A_{260/280}$ value, which should ideally be 1.8. The DNA concentration is determined using a fluorescence-based assay to be as accurate as possible.
23. Barcoded fragments from different fragment generation strategies can be combined as long as each barcode combination is unique. To obtain good coverage for each fragment, it is recommended to pool different samples based on their molarity (n) and not mass (m), as this may bias the resulting sequencing data towards shorter fragments.

24. This procedure can be adapted to any material, the only critical part is to perform an efficient initial PCR to ensure that there is enough template for the subsequent PCR with the primer pairs that attach the barcodes.
25. Ensure that the temperature is maintained at 23°C. Do not exceed 60 minutes of loading, as the Tn5 enzyme will gradually lose activity.
26. To avoid pipetting very small volumes, it is recommended to mix the forward and reverse barcode primers 1:1, resulting in a 5 µM mixture.
27. The gap-filling step at 72°C for 3 minutes is essential to fill the 5' overhangs of the single-stranded linker oligonucleotides to allow for adapter primer binding and amplification of the enzymatically fragmented DNA.
28. Carefully track and document the location of each plasmid sample.
29. This procedure is optimized for acoustic dispensers. This has the advantage of using very small volumes and being tip-free. If an acoustic dispenser is not available, other types of equipment may be suitable. However, care must be taken to avoid cross-contamination of primers with attached barcodes.
30. Instead of pipetting each well into a tube, you can invert the PCR plate into a matching microplate lid and spin the plate briefly to collect all samples in the lid. From there, transfer to a reaction tube of the appropriate size for your amplicon cleanup procedure. **Important:** Test the settings before performing this step with valuable sample material.
31. Settings must be optimized for each incubator to avoid cross-contamination or evaporation, and to achieve optimal aeration and microbial growth.
32. This protocol uses a Singer Instruments Rotor HDA+. Different robots have different types of pinpads and may need to be optimized for this procedure. However, if the sample size is not too large, the procedure can be performed with a manual 384 replica plating stamp.
33. It is recommended to test the primer efficiency on a small scale with the same settings before performing a high throughput experiment.
34. Follow the most recent library preparation protocol for the chemistry provided in the Nanopore Community.
35. We do not use the recommended FFPE enzyme reagent in the outlined in the Oxford Nanopore Technologies protocol for the ligation sequencing protocol.
36. Carefully review the Oxford Nanopore Technologies protocol for the selected library sequencing kit. The kit does not contain all necessary reagents and materials, which must be purchased in advance. E.g. magnetic racks, enzymes and magnetic beads.
37. Wash and store cells according to the most recent procedure provided in the Nanopore Community protocols and return to Oxford Nanopore Technologies as outlined by the company. We do not reuse MinION flow cells. While this option is available, it is recommended that different barcode combinations be used to avoid cross-contamination due to consecutive sequencing experiments.
38. It is recommended to check the Nanopore Community frequently for the latest basecalling software and documentation.

39. It is recommended to refer to the GitHub documentation for the latest user guides and software versions.
40. It is recommended to refer to the Oxford Nanopore Technologies documentation for the latest operating instructions and software versions of Guppy or alternative basecalling software.
41. For the development and deployment of our pipeline, Docker was used to address the complex and diverse dependencies and environmental configurations inherent in bioinformatics tools and applications. This technology ensures consistent, reproducible, and reliable results across multiple operating systems and computing infrastructures.
42. Docker Engine is available on a variety of Linux distributions, MacOS, and Windows OS as a static binary installation. See the DuBA.flow documentation for compatible versions of Docker Desktop.
43. To run Docker on Windows machines, you must have a system running a recent version of Windows 10 or Windows 11 with the Windows Subsystem for Linux 2 (WSL2) installed and enabled. The pipeline was tested using a laptop running Windows 11 Home edition ver. 22H2 with 16 GB of RAM and an intel i5-11300H processor or better.
44. Example files are provided with the Docker image.
45. If a reference is not available or a sequence needs to be assembled *de novo*, the Ref.creator tool (https://github.com/RGSchindler/Ref.creator) can be used to generate a reference or *de novo* assembly based on the sequencing reads. The tool is available as a Docker image on the Docker Hub and its detailed documentation is maintained on GitHub and should be followed.
46. Depending on the Docker settings, you may need to type "`bash`" and press Enter before running the "`python3 DuBaFlow.py`" command.

**Acknowledgments**

This work was supported by the Max Planck Society within the framework of the MaxGENESYS project (DS) and the European Union (NextGenerationEU) via the European Regional Development Fund (ERDF) by the state Hesse within the project "biotechnological production of reactive peptides from waste streams as lead structures for drug development" (DS). We are grateful to all laboratory members for extensive discussions on the development of the DuBA.flow workflow and in particular Tania S. Köbel for her superior technical support. We thank Lars Steinmetz and Kim Remans at EMBL for sharing the transposase expression plasmids from Hennig *et al*., 2018. We thank Tobias Erb and the Erb Lab for constant support with protein purification and unrestricted access to equipment. We thank Ehmad Chehrghani Bozcheloe for visualization of the petri dish and the MinION Nanopore sequencer in Figure 1. All material is available from the corresponding author upon request; except transposase material originates from Hennig *et al*., 2018 and must be requested from the appropriate source.